\documentclass[preprint]{aastex61}

\shorttitle{Eighty-Seven Stars from the NPOI}
\shortauthors{Baines et al.}

\begin{document}

\title{Fundamental Parameters of Eighty-Seven Stars from the Navy Precision Optical Interferometer}

\author{Ellyn K. Baines, J. Thomas Armstrong, Henrique R. Schmitt}
\affil{Naval Research Laboratory, Remote Sensing Division, 4555 Overlook Ave SW, Washington, DC 20375}
\email{ellyn.baines@nrl.navy.mil}

\author{R. T. Zavala, James A. Benson, Donald J. Hutter}
\affil{U.S. Naval Observatory Flagstaff Station, 10391 W. Observatory Rd, Flagstaff, AZ 86001}

\author{Christopher Tycner}
\affil{Central Michigan University, Department of Physics, 1200 S. Franklin St, Mt Pleasant, MI 48859}

\author{Gerard T. van Belle}
\affil{Lowell Observatory, 1400 W. Mars Hill Rd, Flagstaff, AZ 86001} 

\begin{abstract}

We present the fundamental properties of 87 stars based on angular diameter measurements from the Navy Precision Optical Interferometer, 36 of which have not been measured previously using interferometry. Our sample consists of 5 dwarfs, 3 subgiants, 69 giants, 3 bright giants, and 7 supergiants, and span a wide range of spectral classes from B to M. We combined our angular diameters with photometric and distance information from the literature to determine each star's physical radius, effective temperature, bolometric flux, luminosity, mass, and age.

\end{abstract}



\section{Introduction} 

Interferometry is ideally suited to measure the angular diameters of stars, from main-sequence dwarfs \citep[e.g.,][]{2012ApJ...757..112B} to giants \citep[e.g.,][]{2001AandA...377..981W, 2016AJ....152...66B} to supergaints \citep[e.g.,][]{2009MNRAS.394.1925V, 2017AandA...597A...9W} to special cases such as carbon stars \citep{2013ApJ...775...45V}, nearby solar-type stars \citep{2017AandA...597A.137K}, Mira variable stars \citep{2016AandA...587A..12W}, and so on. The direct measurements of these stars' angular diameters are key to determining their fundamental properties such as their physical radii and effective temperatures. These measurements act as a vital check to assumptions inherent in stellar structure and evolution models. 

Interferometric diameters touch many topics of scientific interest. To name a few, they tell us about stars like our Sun and what solar-type stars will become as they evolve \citep[e.g.,][]{2011AandA...526L...4B}. They help determine the ages of stars with imaged companions, so we know whether those companions are older, cooler brown dwarfs or younger, hotter exoplanets \citep[e.g.,][]{2012ApJ...761...57B,2016ApJ...822L...3J}. They act as a direct test of astroseismic relationships \citep[e.g.,][]{2012ApJ...760...32H, 2014ApJ...781...90B}. They characterize exoplanet host stars, which is a vital step in understanding the nature of the companions \citep[e.g.,][]{2012AandA...545A...5L, 2015MNRAS.447..846B}. Furthermore, with the release of $Gaia$ parallaxes, the distance to these targets will be updated and improved \citep{2012AandA...538A..78L}. When their distances are more precisely determined, the stars' physical radii are also more precisely known.

Interferometers have produced excellent survey results in the past, such as in \citet{1999AJ....117..521V}, \citet{1999AJ....118.3032N}, \citet{2003AJ....126.2502M}, and \citet{2013ApJ...771...40B}. This work represents the largest sample of interferometrically measured angular diameters for single stars to date. We present the 87 stars observed with the Navy Precision Optical Interferometer (NPOI) that have diameter measurements with errors of a few percent or, as in most cases, less. This paper in conjunction with the multiplicity study by \citet{2016ApJS..227....4H} highlight the NPOI's utility as a survey instrument. The stars cover most spectral types (B to M) and feature mostly giants (66) with 4 dwarfs, 4 subgiants, 5 bright giants, and 8 supergiants.\footnote{These are based on \emph{SIMBAD} spectral types with preference given to the first luminosity class if a range is given. We update these types later in the paper.} Table 1 lists each star's indentifiers, spectral type, $V$ magnitude, parallax, and metallicity.

The paper is organized as follows: Section 2 discusses the NPOI and our observing process, including the selection and characterization of calibrator stars; Section 3 describes the visibility measurements, the angular diameter dependence on the calculated calibrator diameter, and how we determined various stellar parameters, including the bolometric flux, extinction, luminosity, effective temperature, radius, mass, and age for each target; Section 4 considers what conclusions we can draw from the sample with respect to instrumental performance and previous interferometric measurements (when they exist); and Section 5 summarizes our findings.


\section{Interferometric Observations}

\subsection{The Navy Precision Optical Interferometer}

The NPOI is an optical interferometer located on Anderson Mesa, AZ (see Armstrong et al. 1998 for for the instrument description and Hummel et al. 2003 and Benson, Hummel, \& Mozurkewich 2003 for details about the beam combiner). The NPOI consists of two nested arrays: the four fixed stations of the astrometric array (AC, AE, AW, and AN, which stand for astrometric center, east, west, and north, respectively) that are clustered at the center of the array, and the stations of the imaging array. The latter are arranged along three arms with general north, east, and west orientations. Each arm has ten piers where a siderostat can be installed, which means the imaging array can be reconfigured as needed. 

The NPOI currently has six imaging stations in operation (E3, E6, E7, W4, W7, and N3) and four more will be coming online in the near future (N6, N7, E10, and W10). The stations are labeled according to which arm they are on and how far away they are from the array center, with 1 being closest and 10 being farthest away. The current baselines, i.e., the distance between stations, range from 10 m to 97 m. When the E10 and W10 stations are commissioned, the NPOI will have the longest baseline of any optical interferometer at 432 m. 

The NPOI uses a 12.5-cm diameter region of 50-cm siderostats in both the astrometric and imaging stations. We can combine light from any of the astrometric and imaging stations that are appropriate for our science goals, up to six stations at a time. The current magnitude limit is $\sim$5.5 in the $V$-band under normal conditions $\sim$6.0 in excellent conditions.\footnote{The NPOI will soon receive 3 1-meter telescopes, which will improve the array's sensitivity to $V$ = 9.}

We observed 87 stars from 2004 to 2016, a data set that totals over 100,000 calibrated data points. Table 2 lists the stars observed, the calibrators used, the dates, baselines, and number of observations. We used the ``Classic'' beam combiner that takes data over 16 spectral channels spanning 550 to 850 nm \citep{2003AJ....125.2630H, 2016ApJS..227....4H}. Each observation consisted of a 30-second coherent (on the fringe) scan where the fringe contrast was measured every 2 ms. Every coherent scan was paired with an incoherent (off the fringe) scan that was used to estimate the additive bias affecting fringe measurements \citep{2003AJ....125.2630H}. Scans were taken on one to five baselines simultaneously. Each coherent scan was averaged to a 1-second data point, and then to a single 30-second average. The dispersion of the 1-second data points estimated the internal uncertainties. 

The NPOI's data reduction package $OYSTER$ was developed by C. A. Hummel\footnote{www.eso.org/$\sim$chummel/oyster/oyster.html} and automatically edits data using the method described in \citet{2003AJ....125.2630H}. In addition to that process, we edited individual data points and/or scans that showed large scatter, on the order of 5-$\sigma$ or higher. This was more common in the channels corresponding to the short wavelengths, a long-standing feature in NPOI data, where the channels are narrower, the atmospheric effects are more pronounced, and the avalanche photodiode detectors have lower quantum efficiencies. Removing those short-wavelength scans did not affect the diameter measurements.

The NPOI uses an extensive laser metrology system to measure the three-dimensional motions of the baselines with respect to an Earth-fixed reference system and to determine the absolute wavelength reference \citep[for details, see][]{2003SPIE.4838.1234H}. In order to characterize the stability of the wavelength scale calibration, the NPOI regularly measured the central wavelengths of all the spectrometer channels in a Fourier transform spectrometer mode starting in 2005. The measurements show the central wavelengths are stable with a 0.6 nm (0.1$\%$) scatter \citep{2016ApJS..227....4H}. For data prior to 2005, we incorporated a $\pm$0.5$\%$ error in the wavelength scale. Only five stars include data from 2004, and of those stars only one (HD 172167) used only 2004 data and had an uncertainty $<0.5\%$. We assigned a 0.5$\%$ error to its diameter to account for the uncertainty in the wavelength scale.

\subsection{Selection and Characterization of Calibrator Stars}

The theoretical response of an interferometer for a point source is known. We chose small stars to act as those point-source calibrators, and observed them in sequence with our science target. When we know what the calibrator's visibility should look like, we can compare that to what we see. We corrected for the difference between the theoretical positions and observed data, which is caused mostly by atmospheric effects. 

To estimate the calibrator stars' angular diameters, we created spectral energy distribution (SED) fits based on published $UBVRIJHK$ photometric values obtained from the literature. We used plane-parallel model atmospheres \citep{2004astro.ph..5087C} based on $T_{\rm eff}$, surface gravity (log~$g$), and $E(B-V)$. The stellar models were fit to observed photometry after converting magnitudes to fluxes using \citet{1996AJ....112..307C} for $UBVRI$ and \citet{2003AJ....126.1090C} for $JHK$. See Table 3 for the photometry, effective temperature ($T_{\rm eff}$), log~$g$, and $E(B-V)$ used as well as the resulting angular diameters. This is a relatively simple SED fit, unlike the one described in Section 3.2. For calibrator stars, it is appropriate considering the insensitivity of the final target's angular diameter with regard to the calibrator's diameter (discussed further in Section 3.1). We compared our estimated diameters to those predicted by the \emph{SearchCal} tool provided by JMMC \citep{2016AandA...589A.112C}. The difference between the diameters was an average of only 8$\%$.

We checked every calibrator star for binarity, variability, and rapid rotation. Some of the calibrators chosen featured one or more of those properties, but not to an extent that would affect the calibration process. For the calibrators used here, any binary separation was beyond the detection limit of the configuration used, while the oblateness due to rapid rotation and/or variability in visible wavelengths did not affect the SED fits.

The standard procedure when reducing NPOI data includes smoothing systematic variations in the measured visibilities for the calibrator according to time. We used a smoothing time of 80 minutes, since that was found to be the optimal value for angular diameter measurements as described in \citet{2016ApJS..227....4H}.


\section{Results}

\subsection{Angular Diameter Measurement}

Interferometric diameter measurements use visibility squared ($V^2$). For a point source, $V^2$ is 1 and it is considered completely unresolved. A star is completely resolved when its $V^2$ reaches zero, but naturally a signal of zero is not easily measurable. For a uniformly-illuminated disk, $V^2 = [2 J_1(x) / x]^2$, where $J_1$ is the Bessel function of the first order, $x = \pi B \theta_{\rm UD} \lambda^{-1}$, $B$ is the projected baseline toward the star's position, $\theta_{\rm UD}$ is the apparent uniform disk angular diameter of the star, and $\lambda$ is the effective wavelength of the observation \citep{1992ARAandA..30..457S}. $\theta_{\rm UD}$ results are listed in Table 4. The data are freely available in OIFITS form \citep{2017AandA...597A...8D} upon request.

A more realistic model of a star's disk includes limb-darkening (LD).  If a linear LD coefficient $\mu_\lambda$ is used, then
\begin{eqnarray}
V^2 = \left( {1-\mu_\lambda \over 2} + {\mu_\lambda \over 3} \right)^{-1} 
\times 
\left[(1-\mu_\lambda) {J_1(x_{\rm LD}) \over x_{\rm LD}} + \mu_\lambda {\left( \frac{\pi}{2} \right)^{1/2} \frac{J_{3/2}(x_{\rm LD})}{x_{\rm LD}^{3/2}}} \right]^2 .
\end{eqnarray}
where $x_{\rm LD} = \pi B\theta_{\rm LD}\lambda^{-1}$ \citep{1974MNRAS.167..475H}. We used $T_{\rm eff}$, log $g$ values, and metallicity ([Fe/H]) values from the literature with a microturbulent velocity of 2 km s$^{\rm -1}$ to obtain $\mu_\lambda$ from \citet{2011AandA...529A..75C}. We used the ATLAS model in the \emph{R}-band, since that waveband most closely matched the central wavelength of the NPOI's bandpass. The $T_{\rm eff}$, log $g$, [Fe/H], and $\mu_\lambda$ used and resulting $\theta_{\rm LD}$ measurements are listed in Table 4. Figure 1 shows the $\theta_{\rm LD}$ fit for HD 432 as a representative example. The remaining plots are included in the supplementary material of the \emph{Astronomical Journal}. 

The standard NPOI data reduction sequence incoherently averages the 2~ms data frames to produce 1~s ``points,'' and then averages the points to produce a scan. This two-step averaging procedure is performed separately for each baseline and wavelength channel. In addition to $V^2$, it also yields an estimate of the measurement error for the scan based on the variance of the points within the scan.

However, the uncertainty in a stellar diameter can be significantly underestimated if we feed the $V^2$ and the measurement error estimates into a standard $\chi^2$ minimization routine without regard to the correlations within a scan.  In particular, a calibration error, which can arise naturally because the calibration-star scan is taken at a different time, affects the visibilities for all the baselines and channels within a scan.

To produce an estimate of the diameter uncertainty, we use a modified bootstrap Monte Carlo method devised by \citet{2010SPIE.7734E.103T}, in which we create a large number of synthetic datasets by selecting scans, rather than individual data points, at random.  The width of the distribution of diameters fit to these datasets becomes our measure of the uncertainty in the diameter (see Figure \ref{plot_gauss}). This uncertainty estimate can be as much as an order of magnitude greater than an estimate based only on the within-scan measurement errors.

In order to test the robustness of the calibration process, we changed the calibrator diameter by $\pm$10$\%$ and recalculated target angular diameters. The resulting change in the target's diameters ($\theta_{\rm DIFF}$) was $<1\%$ for 72 stars and between 1$\%$ and 2$\%$ for 12 stars. The remaining stars where $\theta_{\rm DIFF} \geq 3\%$ are:

\emph{HD 109358:} $\theta_{\rm DIFF} = 3\%$, which is larger than the measured angular diameter percent error ($\sigma_{\rm LD}$) of 1.7$\%$. We increased $\sigma_{\rm LD}$ to 3$\%$ to account for $\theta_{\rm DIFF}$.

\emph{HD 120136:} This is the second smallest star in the sample, with $\theta_{\rm LD} = $ 0.822 mas. Its $\sigma_{\rm LD}$ was measured to be 4.6$\%$, which is less than $\theta_{\rm DIFF} = 6\%$. We increased $\sigma_{\theta \rm LD}$ to 6$\%$ to account for $\theta_{\rm DIFF}$.

\emph{HD 120315:} $\theta_{\rm DIFF} = 4\%$, which is far less than $\sigma_{\rm LD} = 15\%$. We left the larger $\sigma_{\rm LD}$ intact.



In most cases (73 out of 87), $\sigma_{\rm LD}$ was larger than $\theta_{\rm DIFF}$.  Of the 14 remaining stars, 12 stars had $\theta_{\rm DIFF} < 1\%$, and the two remaining stars had $\theta_{\rm DIFF} \sim 1\%$. This demonstrated that the assumed calibrator angular diameter of 5$\%$ is reasonable, considering that increasing that uncertainty to 10$\%$ had so little effect on the final angular diameters of the target stars.

\subsection{Stellar Radius, Luminosity and Effective Temperature}

For each target, the parallax from \citet{2007AandA...474..653V}\footnote{The stars presented here were not included in \emph{Gaia's} DR1.} was converted into a distance and combined with our measured diameters to calculate the physical radius $R$. In order to determine each star's luminosity $L$ and $T_{\rm eff}$, we created SED fits using photometric values published in \citet{1965ArA.....3..439L}, \citet{1981PDAO...15..439M}, \citet{1993AandAS..102...89O}, \citet{1990VilOB..85...50J}, \citet{1972VA.....14...13G}, \citet{1970AandAS....1..199H}, \citet{1991TrSht..63....1K}, \citet{1968tcpn.book.....E}, \citet{1966CoLPL...4...99J}, \citet{2003tmc..book.....C}, and \citet{1993cio..book.....G} as well as spectrophotometry from \citet{1983TrSht..53...50G}, \citet{1998yCat.3207....0G}, and \citet{1997yCat.3202....0K} obtained via the interface created by \citet{1997AandAS..124..349M}. The assigned uncertainties for the 2MASS infrared measurements are as reported in \citet{2003tmc..book.....C}, and an uncertainty of 0.05 mag was assigned to the optical measurements. 

We determined the best fit stellar spectral template to the photometry from the flux-calibrated stellar spectral atlas of \citet{1998PASP..110..863P} using the $\chi^2$ minimization technique \citep{1992nrca.book.....P, 2003psa..book.....W}. This gave us the bolometric flux ($F_{\rm BOL}$) for each star and allowed for the calculation of extinction $A_{\rm V}$ with the wavelength-dependent reddening relations of \citet{1989ApJ...345..245C}.

We combined our $F_{\rm BOL}$ values with the stars' distances to estimate $L$ using $L = 4 \pi d^2 F_{\rm BOL}$. We also combined the $F_{\rm BOL}$ with $\theta_{\rm LD}$ to determine each star's effective temperature by inverting the relation,
\begin{equation}
F_{\rm BOL} = {1 \over 4} \theta_{\rm LD}^2 \sigma T_{\rm eff}^4,
\end{equation}
where $\sigma$ is the Stefan-Boltzmann constant and $\theta_{\rm LD}$ is in radians \citep{1999AJ....117..521V}. We follow \citet{2015AandA...582A..49H}, who established a systematic uncertainty of 5$\%$ on their $F_{\rm BOL}$ determinations from a sample of 34 benchmark \emph{Gaia} stars. We therefore assigned an error of 5$\%$ for stars whose SED fits produced errors for $F_{\rm BOL}$ less than 5$\%$. The resulting $R$, $F_{\rm BOL}$, $A_{\rm V}$, $L$, and $T_{\rm eff}$ are listed in Table 5.

Because $\mu_\lambda$ is chosen based on a given $T_{\rm eff}$, we checked to see if $\mu_\lambda$ and therefore $\theta_{\rm LD}$ would change based on our new $T_{\rm eff}$. In most cases, $\mu_\lambda$ changed by an average of 0.01, and the largest difference was 0.11. The resulting $\theta_{\rm LD}$ values changed at most by 1.5$\%$, and the average difference was 0.2$\%$ (0.010 mas). This was well within the uncertainties on $\theta_{\rm LD}$, and re-calculating $T_{\rm eff}$ with the new $\theta_{\rm LD}$ made at most a 47 K difference, which was for the hottest star in the sample (HD 120315, a 0.3$\%$ change), while the average difference was 8 K. These diameters and temperatures all converged after this one iteration, and these are the final values listed in Table 5.

\subsection{Stellar Mass and Age}

To estimate masses and ages for the evolved stars, we used the PARAM stellar model\footnote{http://stev.oapd.inaf.it/cgi-bin/param$\_$1.0} from \citet{2000AandAS..141..371G} with a modified version of the method described in \citet{2006AandA...458..609D} and PARSEC isochrones from \citet{2012MNRAS.427..127B}. For each star, the input parameters were its interferometrically determined $T_{\rm eff}$, its [Fe/H] from the literature, its $V$ magnitude from \citet{Mermilliod}, and its $Hipparcos$ parallax from \citet{2007AandA...474..653V}. The model used these inputs to derive each star's age, mass, radius, $(B-V)_0$, and log~$g$ using the isochrones and a Bayesian estimating method, calculating the probability density function separately for each property in question. \citet{2006AandA...458..609D} qualify mass estimates as ``more uncertain'' than other properties, so the resulting masses listed in Table~6 should be viewed as estimates only.

\section{Discussion}

Several factors can affect a star's visibilities and subsequent angular diameter measurement: variability, binarity, or rapid rotation. None of the stars in our sample are variable to a degree that would be detectable in NPOI data. While some of the stars presented here do have binary companions, we could disregard the secondary star in our angular diameter fits to the primary star due to the separation between the components and/or the magnitude difference between them. \citet{2016ApJS..227....4H} recently demonstrated that the NPOI's detection sensitivity spans 3 to 860 mas with a magnitude difference of 3.0 (for most binary systems) to 3.5 (where the component spectral types differ by less than two). Any companions to our targets were beyond those detection limits.

Three of the stars are rapid rotators with $v \sin i$ higher than 100 km s$^{\rm -1}$: HD 87901 ($\alpha$ Leo, Regulus), HD 159561 ($\alpha$ Oph, Rasalhague), and HD 187642 ($\alpha$ Lyr, Vega). The oblateness for these stars have been measured previously using the CHARA Array for Regulus \citep{2005ApJ...628..439M} and Rasalhague \citep{2009ApJ...701..209Z}, and the NPOI for Vega \citep{2006Natur.440..896P}. We do not directly measure the oblateness of these stars here, since the sampling of the $u-v$ plane for these stars do not give us enough coverage to detect asymmetries.

The size of the data set means we can use it to characterize the NPOI's performance. Figure \ref{error_compare} shows the percent error in $\theta_{\rm LD}$ ($\sigma_{\rm LD}$) versus $\theta_{\rm LD}$. The increase in errors as the diameter approaches 1 mas is expected, considering the resolution limit of the NPOI with the configurations used is $\sim 1$ mas. Above 3.5 mas, the errors are uniformly $\sim 1\%$ or smaller. As the NPOI gets the longer baselines as planned, the limiting resolution will get smaller and the associated errors will decrease. Eighty stars have $\sigma_{\rm LD} \leq 2\%$, which is generally agreed to be the minimal standard of astrophysically useful stellar angular diameter measurements \citep{1997IAUS..189..147B, 2009AandA...501..941H}. Note that one point is left off for the sake of clarity: HD 120315 with a diameter of 0.981 mas and an uncertainty of $\sim 15\%$.


Thirty-six of the 87 the stars presented here do not have previously published interferometric angular diameters (see Table 7). Figure \ref{lit_diam_compare} shows the comparison between the diameters for those stars with published values and our measurements. The match is generally good, with some spread towards the smaller angular diameters that approach the resolution limits of some interferometers. Some of the variations between previous measurements and those presented here may be due to what limb-darkening law was used in the different studies. 

The next step for these stars is to directly measure limb-darkening. Many of the stars presented here have data to or through and beyond the first null, where $V^2$ drops to zero. Before the first null, the visibility curve is dominated by the star's angular diameter. After the first null, second order effects such as limb-darkening become important, and specific limb-darkening models and prescriptions can be directly tested \citep{2001AandA...377..981W}. By verifying which limb-darkening models work the best for most stars, we will know how to characterize stars that are not observable using interferometry.

\section{Summary}

We measured the angular diameters of 87 stars using the NPOI and found good agreement between our measurements and previous measurements when the latter were available. We combined our data with information from the literature to also determine the stars' temperatures, radii, bolometric fluxes, and luminosities. Finally we used the PARAM stellar model to estimate their masses and ages. These diameters will be of special interest when $Gaia$ parallaxes are released with smaller errors than $Hipparcos$ parallaxes, since that will allow us to more precisely measure the stars' physical radii and act as even stricter checks on stellar evolution and structure models. 

\acknowledgments

We thank Brian Mason and William Hartkopf of the U.S. Naval Observatory, Washington, DC for their generosity with regard to data in the NPOI archive. The Navy Precision Optical Interferometer is a joint project of the Naval Research Laboratory and the U.S. Naval Observatory, in cooperation with Lowell Observatory, and is funded by the Office of Naval Research and the Oceanographer of the Navy. This research has made use of the SIMBAD database, operated at CDS, Strasbourg, France. This publication makes use of data products from the Two Micron All Sky Survey, which is a joint project of the University of Massachusetts and the Infrared Processing and Analysis Center/California Institute of Technology, funded by the National Aeronautics and Space Administration and the National Science Foundation. This research has made use of the Jean-Marie Mariotti Center JSDC catalogue, available at http://www.jmmc.fr/catalogue$\_$jsdc.htm.

\clearpage


\startlongtable 
\begin{deluxetable}{ccccccrcc}
\tabletypesize{\scriptsize}
\tablecaption{Sample Star Properties.\label{general_properties}}
\tablehead{
 \colhead{}   & \colhead{}   & \colhead{}    & \colhead{Other} & \colhead{Spectral} & \colhead{$V$}   & \colhead{Parallax} & \colhead{}       & \colhead{}    \\
 \colhead{HD} & \colhead{HR} & \colhead{FK5} & \colhead{Name}  & \colhead{Type}     & \colhead{(mag)} & \colhead{(mas)}    & \colhead{[Fe/H]} & \colhead{Ref} \\ }
\startdata
432    &   21 &    2 & $\beta$ Cas    & F2 III    & 2.27$\pm$0.01 &  59.58$\pm$0.38 &  0.03 & 1 \\
1013   &   45 & 1004 & $\chi$ Peg     & M2 III    & 4.80$\pm$0.01 &   8.86$\pm$0.22 &  0.00 & 2 \\
1522   &   74 &    9 & $\iota$ Cet    & K1 III    & 3.54$\pm$0.02 &  11.88$\pm$0.18 &  0.08 & 1 \\
8512   &  402 &   47 & $\theta$ Cet   & K0 III    & 3.60$\pm$0.01 &  28.66$\pm$0.19 & -0.07 & 1 \\
9927   &  464 &   52 & $\upsilon$ Per & K3 III    & 3.57$\pm$0.01 &  18.41$\pm$0.18 &  0.07 & 3 \\
17709  &  843 & 2198 & 17 Per         & K5.5 III  & 4.54$\pm$0.01 &   6.98$\pm$0.44 & -0.26 & 1 \\
18925  &  915 &  108 & $\gamma$ Per   & G9 III    & 2.93$\pm$0.01 &  13.41$\pm$0.51 & -0.10 & 1 \\
20644  &  999 & 2234 & -              & K3 III    & 4.47$\pm$0.01 &   6.01$\pm$0.25 & -0.27 & 1 \\
\enddata
\tablecomments{Spectral types are from \emph{SIMBAD}, $V$ magnitudes are from \citet{Mermilliod}, parallaxes are from \citet{2007AandA...474..653V}, and [Fe/H] are from the following sources: 1. \citet{2012AstL...38..331A}; 2. no [Fe/H] was available so 0.00 was used; 3. \citet{2015AandA...574A.116R}; 4. \citet{2013AandA...556A.121G}; 5. \citet{2011AandA...525A..90K} and 6. \citet{2016AandA...591A.118S}.}
\end{deluxetable}

\clearpage


\begin{deluxetable}{ccllc}
\tablewidth{0pc}
\tablecaption{Observing Log.\label{observations}}
\tablehead{
 \colhead{Target}  & \colhead{Calibrator} & \colhead{Date} & \colhead{Baselines}      & \colhead{$\#$} \\
 \colhead{HD}      & \colhead{HD}         & \colhead{(UT)} & \colhead{Used$^\dagger$} & \colhead{Obs} \\ }
\startdata
432    & 3360   & 2005 Sep 29 & AW-E6                             & 150 \\
       &        & 2005 Sep 30 & AC-AE, AC-AN, AC-AW, AC-E6, AE-AN, AW-E6 & 600 \\
       &        & 2005 Oct 3  & AC-AN, AE-AN                      &  80 \\
       &        & 2005 Oct 6  & AC-AW, AC-E6, AW-E6               & 120 \\
       &        & 2005 Oct 7  & AC-AN, AC-E6, AE-AN, AW-E6        & 360 \\
       &        & 2005 Oct 8  & AW-E6                             &  89 \\
       &        & 2005 Oct 11 & AW-E6                             & 150 \\
1013   & 886    & 2007 Oct 19 & AC-AE, AC-AN, AC-AW, AE-AN        &  60 \\
       &        & 2007 Oct 20 & AC-AE, AC-E6, AE-AN, AW-E6        & 212 \\
       &        & 2007 Nov 4  & AC-AE, AC-AN, AC-AW, AE-AN        & 156 \\
       & 214923 & 2007 Oct 19 & AC-AE, AW-E6                      &  30 \\
       &        & 2007 Oct 20 & AC-AE, AC-E6, AE-AN, AW-E6        & 212 \\
       &        & 2007 Nov 4  & AC-AE, AC-AN, AC-AW, AE-AN        & 154 \\
1522   & 222603 & 2010 Nov 19 & AC-AE, AC-E6, AE-W7, E6-W7        & 189 \\
       &        & 2010 Dec 8  & AC-AE, AC-W7, AE-W7               &  29 \\
\enddata
\tablecomments{$^\dagger$The maximum baseline lengths are AC-AE 18.9 m, AC-AN 22.8 m, AC-AW 22.2 m, AC-E6 34.4 m, AC-W7 51.3 m, AE-AN 34.9 m, AE-W7 64.2 m, AW-E6 53.3 m, and E6-W7 79.4 m. 
This table shows the information for several stars as an example; the full table is available on the electronic version of the \emph{Astronomical Journal}.}
\end{deluxetable}

\clearpage


\startlongtable 
\begin{longrotatetable}
\begin{deluxetable}{lccccccccccccccc}
\tabletypesize{\scriptsize}
\tablecaption{Calibrator Stars' SED Inputs and Angular Diameters. \label{calibrators}}
\tablehead{
 \colhead{ }  & \colhead{Spec} & \colhead{$U$}   & \colhead{$B$}   & \colhead{$V$}   & \colhead{$R$}   & \colhead{$I$}   & \colhead{$J$}   & \colhead{$H$}   & \colhead{$K$}   & \colhead{$T_{\rm eff}$} & \colhead{log $g$}     & \colhead{}    & \colhead{}         & \colhead{}    & \colhead{$\theta_{\rm est}$} \\
 \colhead{HD} & \colhead{Type} & \colhead{(mag)} & \colhead{(mag)} & \colhead{(mag)} & \colhead{(mag)} & \colhead{(mag)} & \colhead{(mag)} & \colhead{(mag)} & \colhead{(mag)} & \colhead{(K)}           & \colhead{(cm s$^{-2}$)} & \colhead{Ref} & \colhead{$E(B-V)$} & \colhead{Ref} & \colhead{(mas)}              \\ }
\startdata
886    & B2 IV    & 1.75 & 2.61 & 2.83 & 2.88 & 3.06 & 3.50 & 3.64 & 3.77 & 21944 & 3.93 &   1 & 0.02 &  11 & 0.45$\pm$0.02 \\
2905   & B1 I     & 3.50 & 4.30 & 4.16 & 4.12 & 4.09 & 4.14 & 4.15 & 4.01 & 22160 & 2.84 & 2,3 & 0.33 &  12 & 0.36$\pm$0.02 \\
3360   & B2 IV    & 2.62 & 3.47 & 3.66 & 3.74 & 3.92 & 4.14 & 4.25 & 4.25 & 20900 & 3.9  &   4 & 0.03 &   2 & 0.34$\pm$0.02 \\
6658   & A3       & 5.29 & 5.15 & 5.04 & 4.96 & 4.90 & 5.10 & 5.00 & 4.77 &  8511 & 4.04 &   5 & 0.00 & N/A & 0.37$\pm$0.02 \\
11171  & F0 V     & 5.04 & 5.00 & 4.67 & 4.47 & 4.31 & 3.66 & 3.47 & 3.87 &  7244 & 4.24 &   5 & 0.02 &  13 & 0.66$\pm$0.03 \\
11415  & B3 V     & 2.62 & 3.22 & 3.37 & 3.40 & 3.53 & 3.86 & 3.93 & 3.96 & 14250 & 3.38 &   6 & 0.05 &  14 & 0.50$\pm$0.03 \\
17573  & B8 V     & 3.16 & 3.51 & 3.61 & 3.64 & 3.73 & 3.66 & 3.80 & 3.86 & 11749 & 4.14 &   5 & 0.01 &   2 & 0.52$\pm$0.03 \\
25642  & A0 IV    & 4.23 & 4.27 & 4.29 & 4.27 & 4.30 & 4.08 & 4.15 & 4.15 &  9790 &  4.1 &   4 & 0.08 &  13 & 0.50$\pm$0.03 \\
\enddata
\tablecomments{Spectral types are from SIMBAD; $UBV$ values are from \citet{Mermilliod}; $RI$ values are from \citet{2003AJ....125..984M}; $JHK$ values are from \citet{2003tmc..book.....C}; $T_{\rm eff}$, log $g$, and $E(B-V)$ values are from the following sources: 1: \citet{2007astro.ph..3658P}; 2: \citet{2009AandA...501..297Z}; 3: \citet{2000asqu.book.....C} based on SIMBAD spectral type; 4: \citet{2010SPIE.7734E..4EL}; 5: \citet{1999AandA...352..555A}; 6: \citet{2010ApJ...722..605H}; 7: \citet{2016AandA...591A.118S}; 8: \citet{2003AJ....126.2048G}; 9: \citet{2011AandA...525A..71W}; 10: \citet{2001AJ....121.2148G}; 11: \citet{2006MNRAS.371..703S}; 12: \citet{1985ApJS...59..397S}; 13: \citet{1980BICDS..19...61N}; 14: \citet{1992BICDS..40...31F}; 15: \citet{1996AandAS..117..227A}; 16: \citet{1996AandAS..118..481B}; 17: \citet{2010AandA...522A..79V}; 18: \citet{2015MNRAS.454.2863H}; 19: \citet{2005ApJ...626..446R}; 20: \citet{2008ApJS..176..276V}; 21: \citet{2006ApJ...647..312O}; 22: \citet{2012AandA...538A.143K}; 23: \citet{1995yCat.3039....0J}; 24: \citet{2008MNRAS.389.1336K}; 25: \citet{2004AJ....127.1227C}.
$\theta_{\rm est}$ is the estimated angular diameter calculated using the method described in Section 2.}
\end{deluxetable}
\end{longrotatetable}


\startlongtable
\begin{longrotatetable}
\begin{deluxetable}{ccccccccccc}
\tablewidth{0pc}
\tabletypesize{\scriptsize}
\tablecaption{Interferometric Results. \label{inf_results}}

\tablehead{\colhead{Target} & \colhead{$\theta_{\rm UD}$} & \colhead{$T_{\rm eff}$} & \colhead{log $g$}        & \colhead{}    & \colhead{Initial}             & \colhead{$\theta_{\rm LD}$} & \colhead{Final}               & \colhead{$\theta_{\rm LD}$} & \colhead{$\sigma_{\rm LD}$} & \colhead{Max SF} \\ 
           \colhead{HD}     & \colhead{(mas)}             & \colhead{(K)}           & \colhead{(cm s$^{-2}$)} & \colhead{Ref} & \colhead{$\mu_{\rm \lambda}$} & \colhead{(mas)}             & \colhead{$\mu_{\rm \lambda}$} & \colhead{(mas)}             & \colhead{($\%$)} & \colhead{(10$^6$ cycles s$^{\rm -1}$)}               }
\startdata
432    & 2.032 &  6810 & 4.09 &    1 & 0.49 & 2.013$\pm$0.015 & 0.49 & 2.103$\pm$0.015 & 0.7 & 94.9	\\
1013   & 4.041 &  3540 & 0.8  &    2 & 0.83 & 4.367$\pm$0.022 & 0.82 & 4.359$\pm$0.022 & 0.5 & 74.9	\\
1522   & 3.380 &  4390 & 2.0  &    2 & 0.73 & 3.328$\pm$0.062 & 0.70 & 3.310$\pm$0.062 & 1.9 & 119.1	\\
8512   & 2.625 &  4677 & 2.57 &    3 & 0.69 & 2.784$\pm$0.016 & 0.64 & 2.764$\pm$0.016 & 0.6 & 119.2	\\
9927   & 3.389 &  4365 & 2.01 &    3 & 0.73 & 3.649$\pm$0.007 & 0.73 & 3.649$\pm$0.007 & 0.2 & 80.7	\\
17709  & 3.648 &  4050 & 1.7  &    2 & 0.77 & 3.907$\pm$0.021 & 0.77 & 3.907$\pm$0.021 & 0.5 & 81.1	\\
18925  & 3.651 &  5623 & 2.44 &    3 & 0.57 & 3.831$\pm$0.018 & 0.70 & 3.894$\pm$0.018 & 0.5 & 34.3	\\
20644  & 3.476 &  4390 & 2.0  &    2 & 0.71 & 3.680$\pm$0.025 & 0.70 & 3.674$\pm$0.025 & 0.7 & 81.1	\\
\enddata
\tablecomments{The errors on the UD diameter are the same as those on the LD diameter. The initial $\mu_{\lambda}$ is based on the $T_{\rm eff}$ and log $g$ listed in the table, and the final $\mu_{\lambda}$ is based on the new $T_{\rm eff}$ determination. (See Section 3.2 for more details). The $T_{\rm eff}$ and log $g$ come from the following sources: 1. \citet{2007astro.ph..3658P}; 2. \citet{2010SPIE.7734E..4EL}; 3. \citet{1999AandA...352..555A}; 4. \citet{2000asqu.book.....C} based on the star's SIMBAD spectral type; 5. \citet{2003AandA...402..433L}; 6. \citet{2000AJ....119.1448H}; 7. \citet{2011AandA...525A..90K}; 8. \citet{2004ApJS..152..251V}; 9. \citet{2016AandA...591A.118S}; 10. \citet{2011MNRAS.414.1227M}; 11. \citet{2011AandA...531A.165P}; 12. \citet{2002AandA...393..183B}; 13. \citet{2007AandA...475.1003H}; and 14. \citet{2010AN....331..349H}. Max SF is the maximum spatial frequency for that star's diameter measurement.}
\end{deluxetable}
\end{longrotatetable}

\clearpage


\startlongtable 
\begin{longrotatetable} 
\begin{deluxetable}{ccccccccc}
\tablewidth{0pc}
\tabletypesize{\scriptsize}
\tablecaption{Derived Stellar Parameters. \label{derived_results}}

\tablehead{\colhead{Target} & \colhead{Spectral}       & \colhead{$R$}         & \colhead{$\sigma_R$} & \colhead{$F_{\rm BOL}$}                      & \colhead{$A_{\rm V}$}           & \colhead{$T_{\rm eff}$} & \colhead{$\sigma_T$} & \colhead{$L$}        \\ 
           \colhead{HD}     & \colhead{Type} & \colhead{($R_\odot$)} & \colhead{($\%$)}     & \colhead{(10$^{-6}$ erg s$^{-1}$ cm$^{-2}$)} & \colhead{(mag)} & \colhead{(K)}           & \colhead{($\%$)}        & \colhead{($L_\odot$)} }
\startdata
432    & F2 III   &   3.79$\pm$0.04                       & 1.0  &  3.04$\pm$0.15 & 0.00$\pm$0.01 &  6739$\pm$88   & 1.3 & 26.8$\pm$1.4	\\
1013   & M2 III   &  52.88$\pm$$^{\rm 1.37}_{\rm 1.31}$   & 2.6  &  1.38$\pm$0.07 & 0.31$\pm$0.01 &  3842$\pm$49   & 1.3 & 549.9$\pm$38.8	\\
1522   & K1 III   &  29.95$\pm$$^{\rm 0.72}_{\rm 0.73}$   & 2.4  &  1.70$\pm$0.09 & 0.39$\pm$0.01 &  4645$\pm$73   & 1.6 & 376.8$\pm$22.0	\\
8512   & K0 III   &  10.37$\pm$0.09                       & 0.9  &  1.53$\pm$0.08 & 0.28$\pm$0.02 &  4951$\pm$64   & 1.3 & 58.3$\pm$3.0	\\
9927   & K3 III   &  21.30$\pm$0.21                       & 1.0  &  1.54$\pm$0.08 & 0.00$\pm$0.01 &  4316$\pm$54   & 1.3 & 142.1$\pm$7.6	\\
17709  & K5 III   &  60.16$\pm$$^{\rm 3.58}_{\rm 4.06}$   & 6.7  &  1.32$\pm$0.07 & 0.40$\pm$0.02 &  4014$\pm$51   & 1.3 & 847.5$\pm$114.9	\\
18925  & G8 III   &  31.21$\pm$$^{\rm 1.15}_{\rm 1.24}$   & 4.0  &  2.24$\pm$0.11 & 0.00$\pm$0.02 &  4589$\pm$58   & 1.3 & 389.6$\pm$35.5	\\
20644  & K3 III   &  65.70$\pm$$^{\rm 2.66}_{\rm 2.89}$   & 4.4  &  1.82$\pm$0.17 & 1.13$\pm$0.02 &  4485$\pm$108  & 2.4 & 1576$\pm$199	\\
\enddata
\tablecomments{The spectral types are those that provide the best SED fit as described in Section 3.2. The SED fits are also the source of $F_{\rm BOL}$ and $A_{\rm V}$. The other parameters are derived as described in Section 3.2.
Note 1: The large error on the parallax measurement made the radius determination unuseful, considering the error would be more than 50$\%$.
Note 2: Because the errors on the parallax measurements are unusually high ($\geq$20$\%$), the errors on the subsequent luminosity calculations render them unhelpful for these stars.}
\end{deluxetable}
\end{longrotatetable}

\clearpage


\startlongtable
\begin{deluxetable*}{ccc}
\tablecaption{Stellar Masses and Ages. \label{mass_age}}
\tablehead{\colhead{Target} & \colhead{Mass}        & \colhead{Age} \\ 
           \colhead{HD}     & \colhead{($M_\odot$)} & \colhead{(Gyr)}}
\startdata
432    & 2.02$\pm$0.03 & 1.00$\pm$0.05	\\
1013   & 1.06$\pm$0.11 & 7.70$\pm$2.01	\\
1522   & 3.70$\pm$0.10 & 0.23$\pm$0.03	\\
8512   & 2.06$\pm$0.14 & 1.08$\pm$0.23	\\
9927   & 1.75$\pm$0.15 & 1.70$\pm$0.40	\\
17709  & 1.34$\pm$0.18 & 3.10$\pm$1.20	\\
18925  & 3.64$\pm$0.25 & 0.23$\pm$0.05	\\
20644  & 3.07$\pm$0.39 & 0.35$\pm$0.14	\\
\enddata
\tablecomments{Masses and ages were determined using the PARAM stellar model described in Section 3.3. An error of 0.05 was assigned to [Fe/H] (a model input), which are from the sources listed in Table 1.}
\end{deluxetable*}

\clearpage


\startlongtable
\begin{deluxetable*}{cccl}
\tabletypesize{\scriptsize}
\tablecaption{Interferometric Angular Diameter Comparison. \label{lit}}

\tablehead{\colhead{Target} & \colhead{$\theta_{\rm LD,here}$} & \colhead{$\theta_{\rm LD,previous}$} & \colhead{ }      \\ 
           \colhead{HD}     & \colhead{(mas)}                       & \colhead{(mas)}                      & \colhead{Reference} } 
\startdata
432    & 2.103$\pm$0.015 & 2.218$\pm$0.058 & \citet{2016ApJS..227....4H}   \\ 
       &                 & 1.698$\pm$0.023 & \citet{2011ApJ...732...68C}   \\
       &                 &  2.12$\pm$0.05  & \citet{1999AJ....118.3032N}   \\
1013   & 4.359$\pm$0.022 & N/A             & N/A                           \\
1522   & 3.310$\pm$0.062 & 3.325$\pm$0.010 & \citet{2009MNRAS.399..399R}   \\
8512   & 2.764$\pm$0.016 & 2.817$\pm$0.044 & \citet{2009MNRAS.399..399R}   \\
9927   & 3.649$\pm$0.007 &  3.76$\pm$0.07  & \citet{1999AJ....118.3032N}   \\
17709  & 3.907$\pm$0.021 & 4.056$\pm$0.041 & \citet{2003AJ....126.2502M}   \\
\enddata
\tablecomments{$^\ast$No LD diameter was provided, therefore we list the UD diameter here. Figure 4 shows a graphical representation of this table. If more than one diameter was available in the literature, we used the most recent one when plotting the results in Figure \ref{lit_diam_compare}.}
\end{deluxetable*}

\clearpage


\begin{figure}[h]
\includegraphics[width=1.0\textwidth]{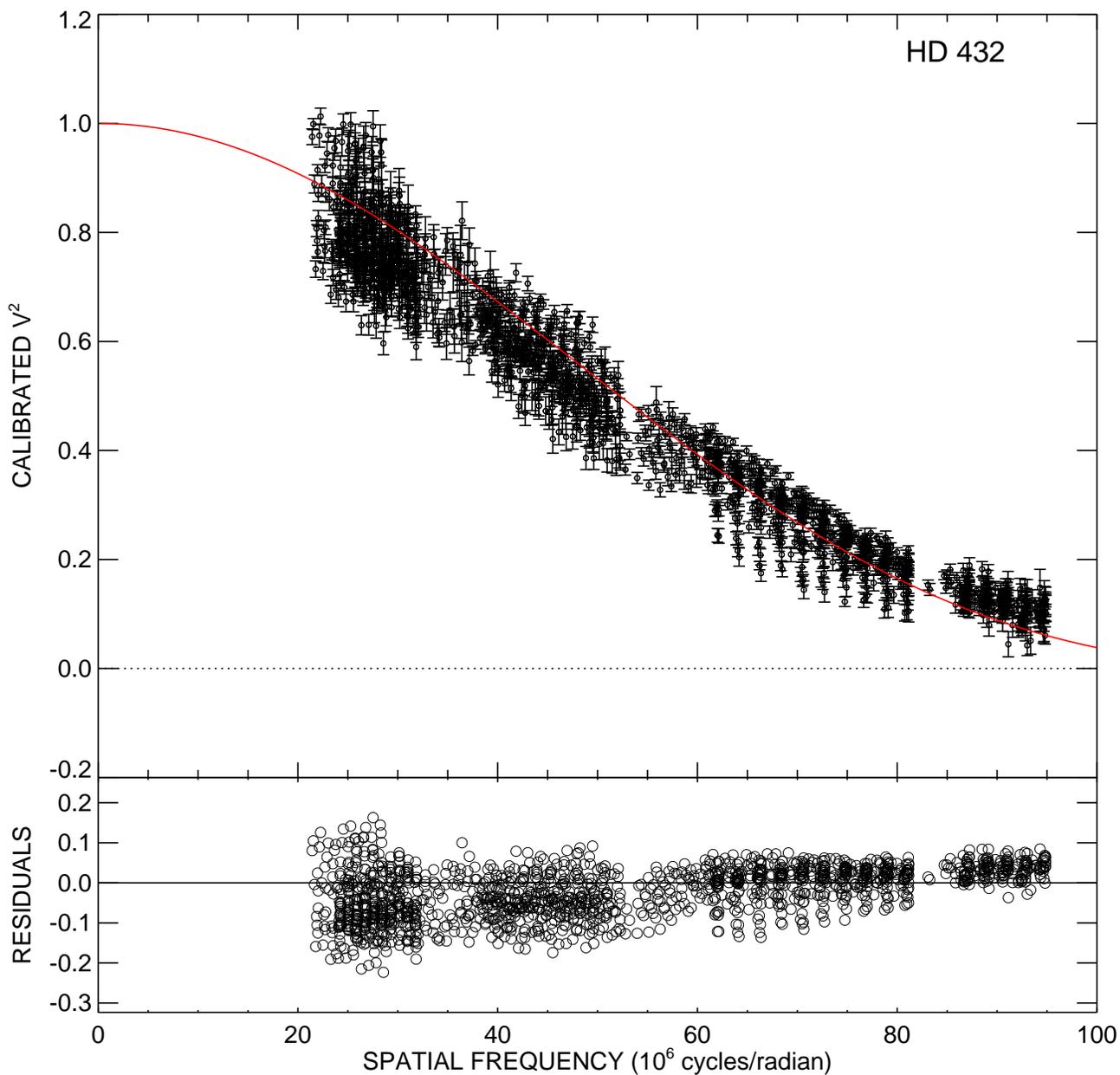}
\caption{\emph{Top panel:} The $\theta_{\rm LD}$ fit for HD 432. The solid lines represent the visibility curve for the best fit $\theta_{\rm LD}$, the points are the calibrated visibilities, and the vertical lines are the measurement uncertainties. \emph{Bottom panel:} The residuals (O-C) of the diameter fit to the visibilities. The plots for the remaining stars are available on the electronic version of the $Astronomical Journal$.}
  \label{visvsfreq1}
\end{figure}

\clearpage

\begin{figure}[h]
\includegraphics[width=1.0\textwidth]{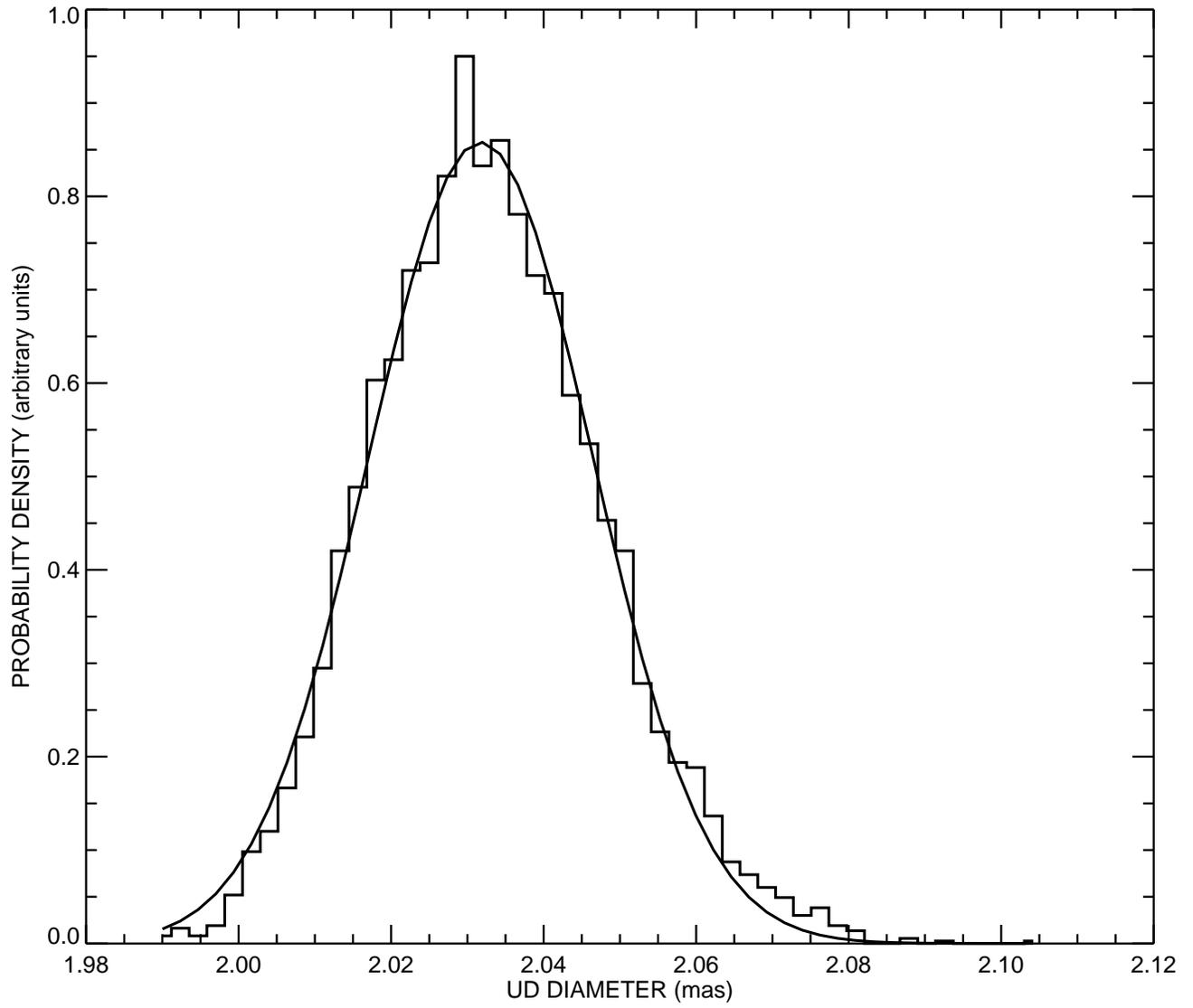}
\caption{An example probability density solution for the diameter fit to HD 432 visibilities as described in Section 3.1.}
  \label{plot_gauss}
\end{figure}

\clearpage

\begin{figure}[h]
\includegraphics[width=1.0\textwidth]{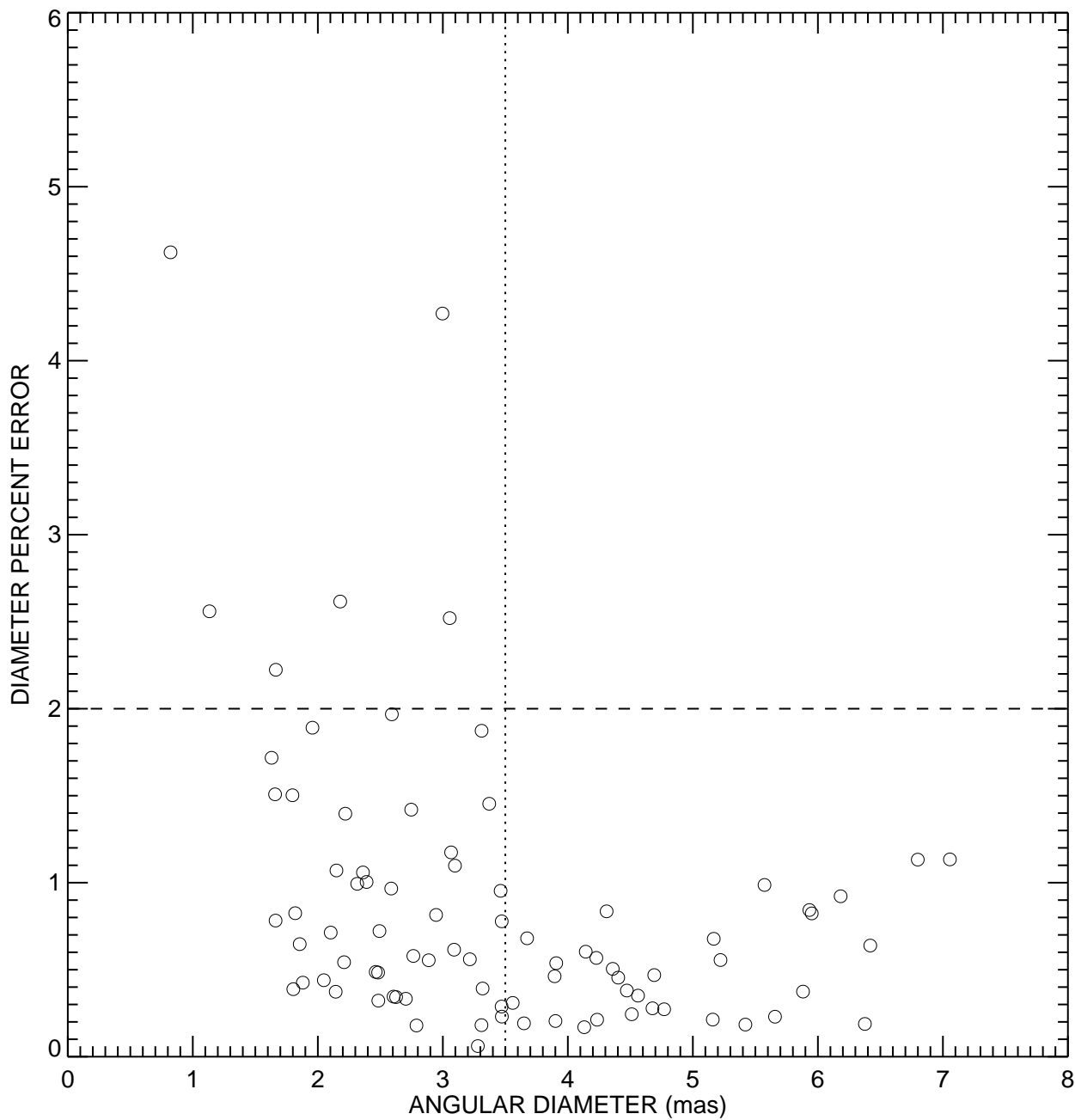}
\caption{Characterizing the NPOI performance based on the percent error in the limb-darkened diameter measurement ($\sigma_{\rm LD}$) versus $\theta_{\rm LD}$. The horizontal dashed line shows the $\sigma_{\rm LD}$=2$\%$ cut-off that is the minimal standard of astrophysically useful measurements, while the vertical dotted line shows the 3.5-mas cut-off where $\sigma_{\rm LD}$ errors are $\sim 1 \%$ or better. The star with the highest error (HD 120315, $\sigma_{\rm LD} =15\%$) is not included with this plot so that the spread of the other points is more easily visible.}
  \label{error_compare}
\end{figure}

\clearpage

\begin{figure}[h]
\includegraphics[width=1.0\textwidth]{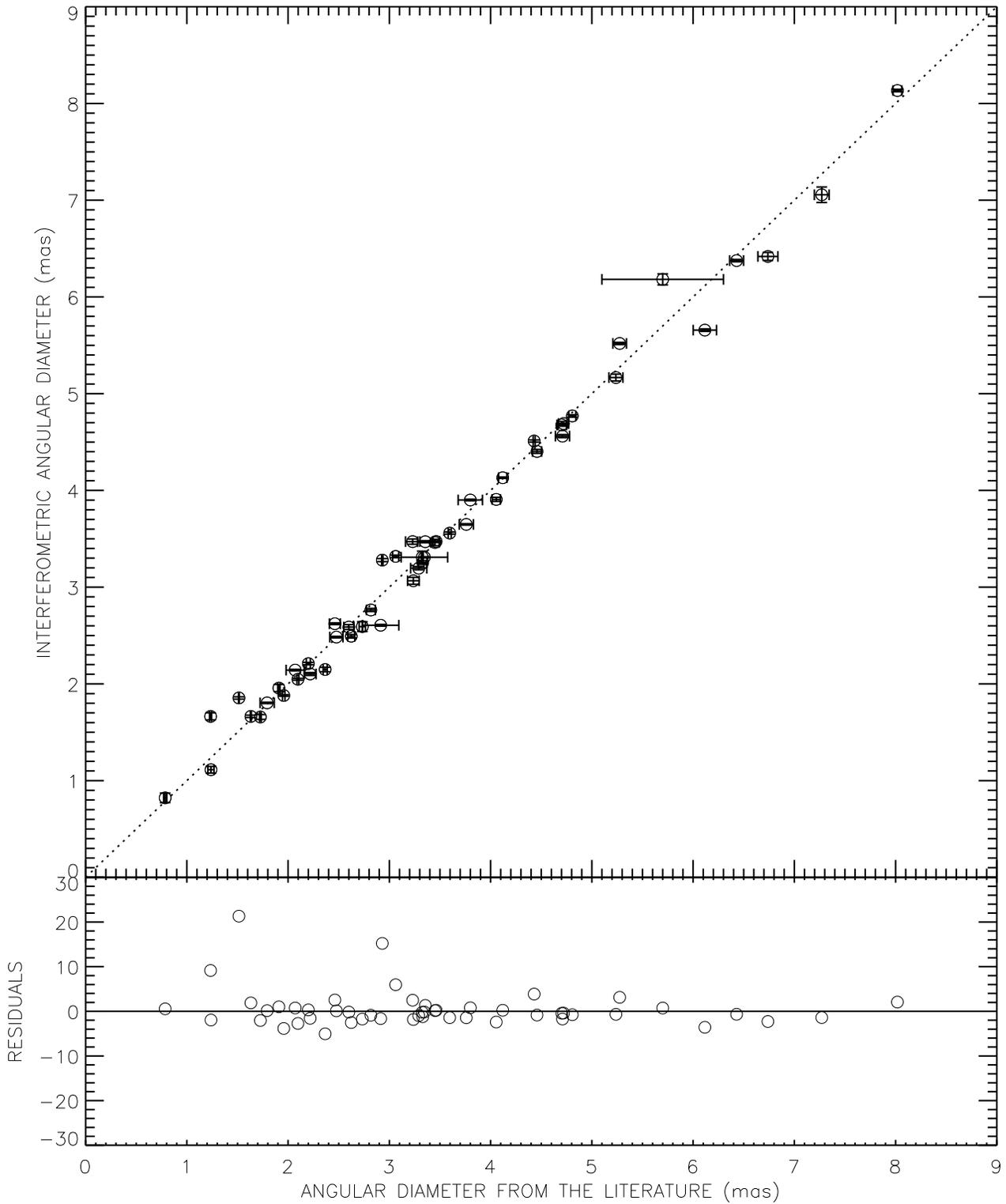}
\caption{\emph{Top panel:} Comparison of interferometrically measured angular diameters versus diameters from the literature. The error bars for the interferometric diameters are often smaller than the circle that indicates that measurement. The dotted line is the 1:1 ratio. When more than one measurement was available in the literature, we used the most recent measurement (see Table 7). \emph{Bottom panel:} The residuals were calculated as follows: ($\theta_{\rm interferometry} - \theta_{\rm literature})$ $\times$ (combined error)$^{-1}$.
}
  \label{lit_diam_compare}
\end{figure}

\clearpage

\end{document}